# Diffusion in a two Phases System: Application to Ion Exchange in Silicate Glasses

Guglielmo Macrelli[1]

[1]*Isoclima SpA – R&D Department*

*Via A.Volta 14, 35042 Este (PD), Italy*

*guglielmomacrelli@hotmail.com*

[*]*Correspondence: guglielmomacrelli@hotmail.com*

**Abstract**

Diffusion in a two phases system is a classical problem discussed in the literature. The general solution of the one-dimensional case to this problem is revisited and a detailed derivation is proposed. The solution is discussed in relationship with ion exchange in silicate glasses. The general solution is compared with the complementary error function (*erfc*) solution and the assumptions under which the general solution reduces to the *erfc* are identified. Interface or glass surface issues related to ion exchange are discussed in terms of ratio between the diffusion coefficients of the diffusing alkali ions in the glass and in the ion reservoir phases respectively. Another relevant issue can be identified in the assumption of instantaneous equilibrium at the two phases interface once the contact is established. Even though this last assumption is somehow unphysical it is the common approach used in technological application of ion exchange in silicate glass which is justified when the contact time (immersion time) of glass with ion reservoir is much larger than the time to achieve thermodynamic equilibrium at the surface.

1. **Introduction**

Ion Exchange of ions *B* in silicate glasses for ions *A* in a reservoir is usually treated as an interdiffusion problem in glass[1,2,3,4,5]. Generally, issues related to kinetic in the ion reservoir and to the interface between glass and the ion reservoir are not systematically treated. This approach is justified based on the assumptions that the source of diffusing ions (*A* ions) works as an infinite reservoir of ions with no kinetic restrictions for the transport of ions within the reservoir and, additionally, a thermodynamic equilibrium is almost instantaneously achieved at the interface between the ion reservoir and glass surface. The above assumptions allow to establish a boundary condition of constant concentration, $c_{sA}$ of the diffusing (incoming) ions *A* at the glass surface. Considering a one-dimensional problem and a constant interdiffusion coefficient $D_{AB}$, the solution to the diffusion equation for the concentration of the diffusion ions in glass $c_A(x,t)$ can be identified in the well-known complementary error function[3,4]

$$c_A(x,t) = c_{sA} \cdot erfc\left(\frac{x}{2\sqrt{D_{AB} \cdot t}}\right). \quad (1)$$

In equation (1), *x* is the spatial coordinate while *t* is the time. The reservoir/glass interface thermodynamic equilibrium condition has been extensively discussed elsewhere[6] while the kinetics of ions *A* in the reservoir and its influence on the kinetics of ions in the glass has been discussed, in the framework of ion exchange, by Schaeffer and Heinze[7]. The solution proposed by Schaeffer and Heinze is coming from previous studies of Oel and Jost[8] and Jost[9]. One purpose of the present study is to provide a detailed, step by step, derivation of this solution which is presented in Appendix 1. The most popular ion sources used in technological applications of in ion exchange in silicate glasses are molten nitrate salts. Transport properties, namely diffusion coefficients, of cation ions (usually alkali ions) are not usually reported in ion exchange studies. A convenient source of data about transport properties of molten salts can be identified in the literature[10]. The present study fucuses attention on the problem of the effect of kinetics in the reservoir phase and on the consequences to the interface (*aka*: "glass surface"). When the ions diffusion coefficient in the ion reservoir phase is, for any reason, comparable with the one in the glass phase the effect is a reduction of concentration at glass surface. This has an evident effect on the boundary condition limit at glass surface. It will be additionally stressed how the assumption of "instantaneous equilibrium condition at reservoir/glass interface" is crucial to accept the constant surface concentration boundary condition leading to the "erfc" type concentration solution (1).

## 2. The solution to the mathematical problem.

Ion Exchange is defined as a binary chemical reaction between two subsystems, identified as 1 and 2, in mutual contact (See Figure 1). Ion exchange in silicate glasses can be represented as a two phases diffusion problem that can be schematically represented as depicted in Figure 1. There are indicated the two phases: phase 1 is the ions reservoir while phase 2 is the glass matrix.

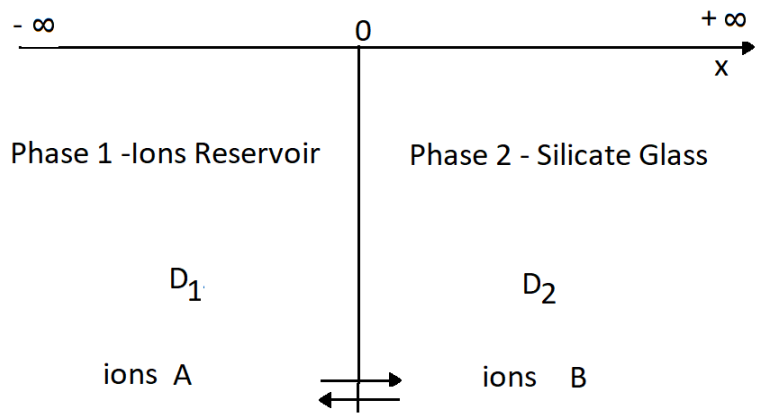

**Figure 1 – Diffusion in a two phases system, $D_1$ and $D_2$ are the diffusion coefficients of ions $A$ in the first and second phases.**

Equations and boundary conditions for the two phases diffusion problem are reported in the following equations numbered from (2) (Diffusion Equations) to (3) (initial condition) and (4) and (5) (Boundary conditions). $C_1(x,t)$ and $C_2(x,t)$ are the concentrations of ions $A$ respectively in phase 1 and phase 2. Initial condition (3) is the condition of constant concentration respectively in phase 1 ($C_I$) and phase 2 ($C_{II}$). Condition (4) is a zero-flux condition at infinite while conditions (5) are interface boundary conditions, respectively: a continuity concentration condition (Named Nerst condition by Schaeffer and Heinze[7]) and a continuity flux condition at the two phases interface.

$$x < 0 \, , \, \frac{\partial C_1}{\partial t} = D_1 \frac{\partial^2 C_1}{\partial x^2} ; \quad x > 0 \, , \, \frac{\partial C_2}{\partial t} = D_2 \frac{\partial^2 C_2}{\partial x^2} , \qquad (2)$$

$$t = 0 ; \, x < 0 , C_1(x,0) = C_I ; \, x > 0 , C_2(x,0) = C_{II} \quad , \qquad (3)$$

$$x \rightarrow -\infty , \, \frac{\partial C_1}{\partial x} = 0 \; ; \; x \rightarrow \infty , \, \frac{\partial C_2}{\partial x} = 0 , \qquad (4)$$

$$t > 0 \;;\; x = 0 \;,\; C_2(0,t) = \chi C_1(0,t) \;\;;\;\; -D_1 \left.\frac{\partial C_1}{\partial x}\right|_{x=0} = -D_2 \left.\frac{\partial C_2}{\partial x}\right|_{x=0}. \tag{5}$$

It is important to stress that condition (5) works when $t > 0$. The continuity between conditions (3) and the first condition (5) is surely a problematic issue to be considered from the purely mathematical point of view. Condition $C_2(0,t) = \chi C_1(0,t)$ represents the situation of an almost instantaneous equilibrium condition achieved by the system once the reservoir and glass are put in contact. In a realistic situation it is expected that the $\chi$ factor is not constant but depends somehow on time, $\chi(t)$ allowing a more physical condition where surface equilibrium is achieved in a small but finite time $\tau$. The assumption we are making here is to consider such time $\tau$ small enough to consider $\chi$ a constant. Since we are dealing with a problem where boundary conditions are not time depending, the solution to this problem can be conveniently approached with the following "ansatz"[7,8,9] for both $C_1(x,t)$ and $C_2(x,t)$:

$$C_i(x,t) = E + F \cdot erf\left(\frac{x}{2\sqrt{D_i t}}\right) \;;\; i = 1,2 \tag{6}$$

The solution for $C_1(x,t)$ and $C_2(x,t)$, that is concentration of ions $A$ in phase 1 and phase 2 respectively, reads:

$$C_1(x,t) = C_I + \frac{\sqrt{D_2}\left(C_{II} - \chi C_I\right)}{\chi\sqrt{D_2} + \sqrt{D_1}} \cdot \left[1 + erf\left(\frac{x}{2\sqrt{D_1 t}}\right)\right] \tag{7i}$$

$$C_2(x,t) = C_{II} - \frac{\sqrt{D_1}\left(C_{II} - \chi C_I\right)}{\chi\sqrt{D_2} + \sqrt{D_1}} \cdot \left[erfc\left(\frac{x}{2\sqrt{D_2 t}}\right)\right], \tag{7ii}$$

(7ii) is the solution reported in Schaeffer and Heinze[7]. The justification of equations (7i) and (7ii) is not reported by Schaeffer and Heinze[7], additionally the Jost and Oel study[8] and the book of Jost[9] do not provide a full treatment leading to solution (7). The justification of solution (7) through "ansatz" (6) is straightforward even if not trivial hence, a detailed justification is presented in Appendix 1. A convenient way to express solution (7ii) is by dividing the numerator and denominator of the second term by the square root of $D_1$.

$$C_2(x,t) - C_{II} = -\frac{(C_{II} - \chi C_I)}{\chi\sqrt{\frac{D_2}{D_1}} + 1} \cdot \left[ erfc\left(\frac{x}{2\sqrt{D_2 t}}\right) \right] = \frac{(\chi C_I - C_{II})}{\chi\sqrt{\frac{D_2}{D_1}} + 1} \cdot \left[ erfc\left(\frac{x}{2\sqrt{D_2 t}}\right) \right], \qquad (8)$$

defining:

$$\delta_R = \frac{1}{\chi\sqrt{\frac{D_2}{D_1}} + 1}, \qquad (9)$$

the solution finally reads:

$$C_2(x,t) - C_{II} = \delta_R (\chi C_I - C_{II}) \cdot \left[ erfc\left(\frac{x}{2\sqrt{D_2 t}}\right) \right]. \qquad (10)$$

The following inequality can be set:

$$0 \leq \delta_R \leq 1. \qquad (11)$$

The concentration in phase 2 can be expressed as excess of concentration ($Ca$) in respect of the initial constant value $C_{II}$. Equation (10) results:

$$Ca(x,t) = \delta_R C_S \cdot \left[ erfc\left(\frac{x}{2\sqrt{D_2 t}}\right) \right]; \quad C_S = \chi C_I - C_{II}. \qquad (12)$$

At a first glance, apart from the $\delta_R$ term, equation (12) looks very similar to equation (1). In the following tables ( I and III) and figures ( Figure 2 and Figure 3) it is reported a parametric evaluation of equation (12) for the concentration of diffusing ions in phase II. The curves are calculated considering a total exposure time of 24 hours ($t = 8.64 \cdot 10^4$ s) and $C_S = 1$.

Table I – Calculation of $Ca(x,t)$ in four cases of diffusion coefficient in phase 1. Diffusion coefficient of phase 2 - $D2 = 1 \cdot 10^{-10}$ cm$^2$/s; $\chi = 0.5$.

| Function | $D_1$(cm$^2$/s) | Ca(0) |
|---|---|---|
| Ca1 (x,t) | $1 \cdot 10^{-5}$ | 0.499 |
| Ca2 (x,t) | $1 \cdot 10^{-8}$ | 0.476 |
| Ca3(x,t) | $1 \cdot 10^{-9}$ | 0.432 |
| Ca4 (x,t) | $1 \cdot 10^{-10}$ | 0.333 |

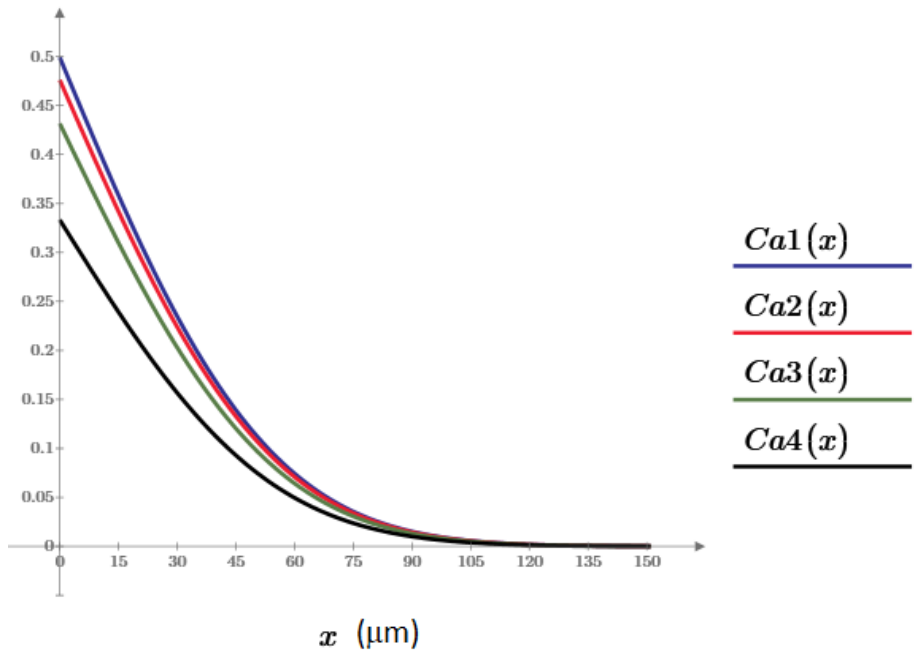

**Figure 2 – Case study of Table I. The $x$ coordinate is in (μm).**

Table II– Calculation of $Ca(x,t)$ in four cases of diffusion coefficient in phase 1. Diffusion coefficient of phase 2 - $D2 = 1 \cdot 10^{-10}$ cm$^2$/s; $\chi = 1.0$.

| Function | $D_1$(cm$^2$/s) | Ca(0) |
|---|---|---|
| Ca1 (x,t) | $1 \cdot 10^{-5}$ | 0.997 |
| Ca2 (x,t) | $1 \cdot 10^{-8}$ | 0.909 |
| Ca3(x,t) | $1 \cdot 10^{-9}$ | 0.760 |
| Ca4 (x,t) | $1 \cdot 10^{-10}$ | 0.500 |

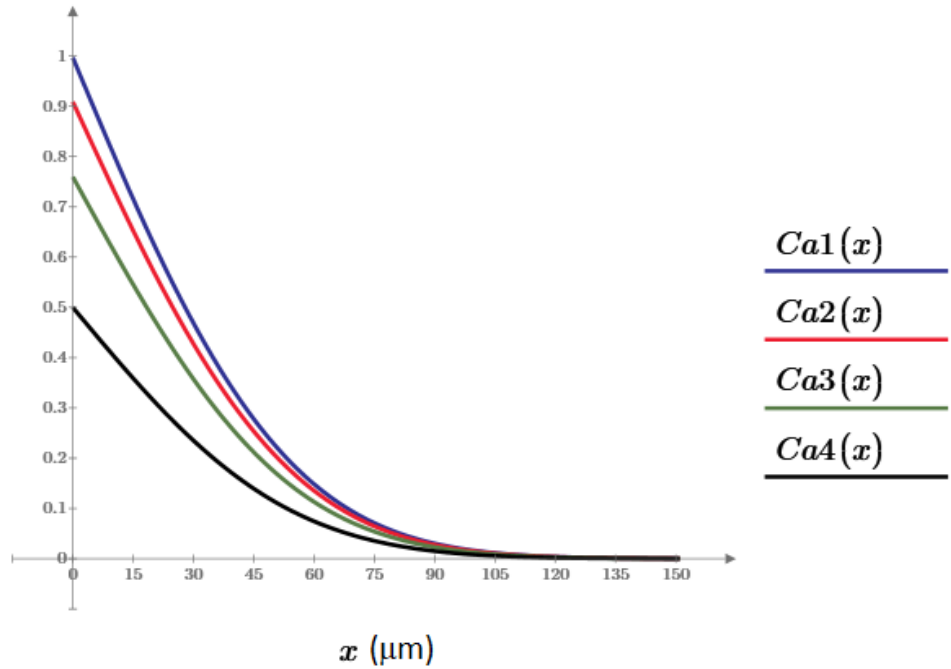

**Figure 3 – Case study of Table II. The *x* coordinate is in (μm).**

In the following figure 4 it is reported the evaluation of equation (12) for the concentration of diffusing ions in phase 2 with fixed diffusion coefficients in the two phases but changing the $\chi$ parameter. The curves are calculated according to values of Table III, considering a total exposure time of 24 hours (t = 8.64·10$^4$ s) and assuming $C_S = 1$.

Table III– Calculation of *Ca(x,t)* in in phase 2 for four cases of parameter $\chi$. Diffusion coefficient of phase 2 - $D_2 = 1·10^{-10}$ cm$^2$/s, Diffusion coefficient of phase 1 – $D_1 = 1·10^{-5}$ cm$^2$/s,

| Function | $\chi$ | Ca(0) |
|---|---|---|
| Ca1 (x,t) | 1.0 | 0.997 |
| Ca2 (x,t) | 0.8 | 0.798 |
| Ca3(x,t) | 0.6 | 0.599 |
| Ca4 (x,t) | 0.4 | 0.399 |

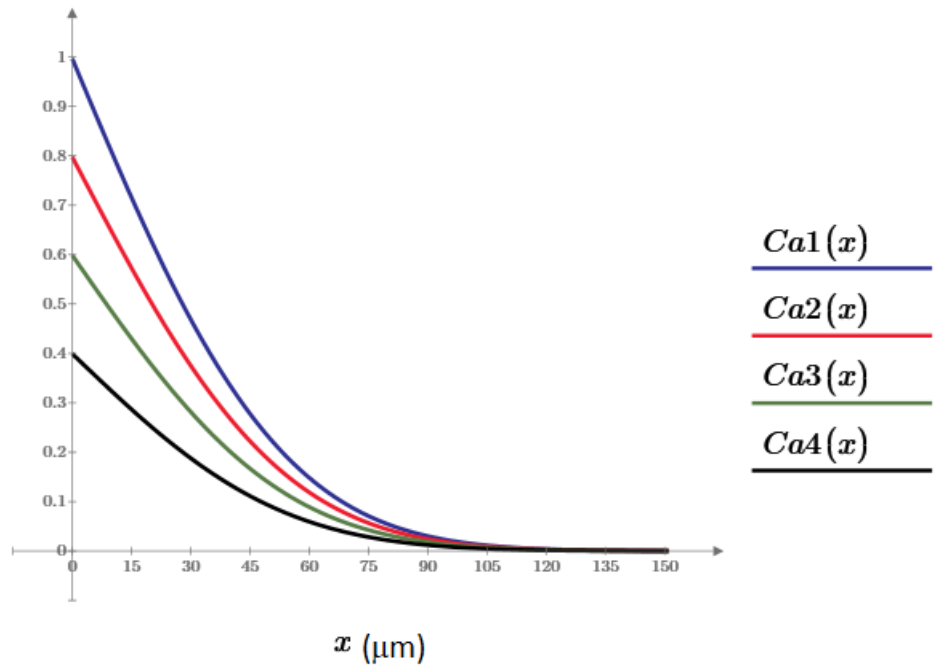

**Figure 4** – Case study of Table III. The *x* coordinate is in (μm).

Tables I and II and the corresponding figures, Figure 2 and Figure 3, represent the effect of the kinetics in the reservoir phase (phase I) to the resulting excess of concentration $C_a(x,t)$ in the glass phase (Phase 2) while Table III and Figure 4 represent the effect to the concentration in the glass phase of the interfacial coefficient $\chi$, with a fixed kinetic transport in the reservoir (Fixed both $D_1$ and $D_2$).

### 3. Application to Ion Exchange in silicate glasses

The application of solution (7) to ion exchange in silicate glasses has been discussed in the literature by Schaeffer and Heinze[7]. In ion exchange application to silicate glasses, the ion reservoir (Phase 1) is a molten alkali nitrates salt that can be either a mass of molten salt in a container or a composite slurry or gel-type coating made of the molten alkali nitrate hold in position on the glass surface by a suitable medium. In order to have ion exchange the glass shall contain a certain amount of an alkali (*B* ions) oxide where the alkali in the glass is to be exchanged for the other alkali (*A* ions) in the molten salt medium. The ion exchange reaction can be represented by equation (13):

$$A + \bar{B} \rightleftarrows B + \bar{A}, \tag{13}$$

where A and B are the ions in the ion reservoir 1 and $\bar{A}$ and $\bar{B}$ are the same ions in the glass. The substitution of alkali ions in the glass matrix with larger alkali ions coming from the molten salt generates residual stress (if the process is performed below the glass transformation range) which is technologically used to increase glass strength[1,2,3,4]. Additionally, because of the different polarizabilities of the ions, ion exchange modifies the glass refractive index allowing an optical waveguide effect on the glass surface[2,3,11]. Typically, ion exchange in silicate glass is performed by immersing the glass articles in a bath of molten nitrates. The main assumption to be considered is that ion exchange is a binary process where the total concentration of alkali is conserved in both phases: ion reservoir (phase 1) and glass (phase 2). The second assumption is that the ion reservoir is so large in comparison with glass that it works as a "reservoir" in the sense that its alkali concentration is only marginally affected by the alkali coming from glass phase. It has been recently stressed in the literature[12] how the excess of original alkali (*A*) in the ion reservoir is crucial to maintain the efficiency of the ion exchange process. This last statement has also been put in evidence by studies about the polluting effect[6,13] of an excess of alkali coming from the glass (*B*) in the ion reservoir. The diffusion coefficient of the alkali ions in the molten salt is typically in the range of *$1 \cdot 10^{-5} cm^2/s < D_1$ $<5 \cdot 10^{-5} cm^2/s$* as reported in Figure 5 based on data of Sundheim[10]. The above considerations lead to the identification of the possible limiting factors for ion exchange in silicate glasses:

- ❖ Kinetics of A and B ions in the molten salts, in our case represented by the diffusion coefficient $D_1$.
- ❖ Saturation limit of the B ion sites available on glass surface.

The second limiting factor is due to the binary character of the ion exchange phenomenon herewith discussed that can be expressed in terms of the sum of concentrations of ions A and B that shall be constant: *[A]+[B] = constant*.

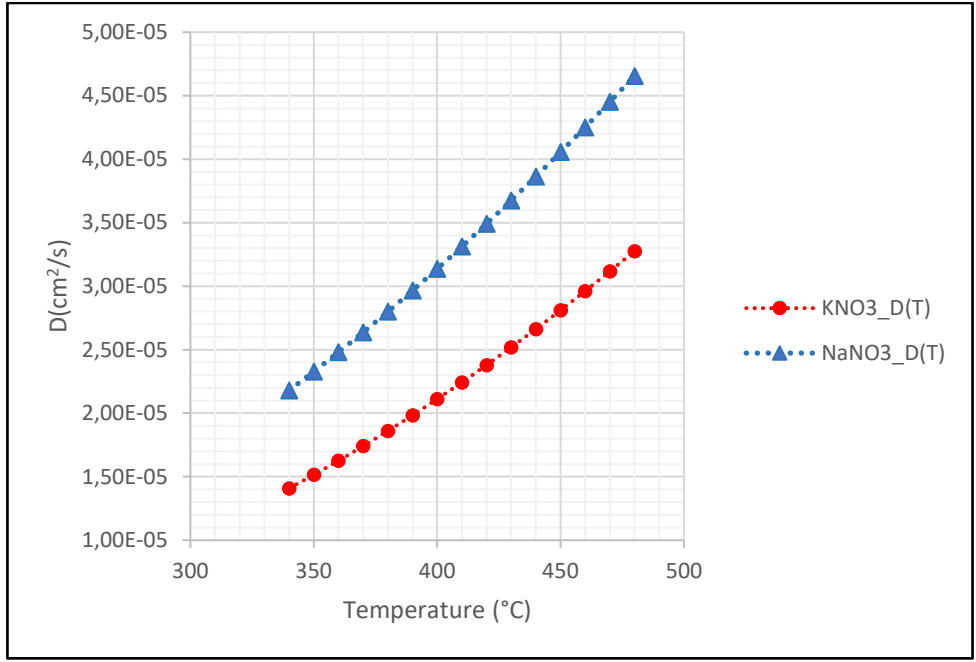

**Figure 5 – Diffusion coefficient of alkali ions (K+ in KNO₃ and Na+ in NaNO₃) as a function of temperature of the molten salt. Data from Sundheim[10].**

A first consideration can be proposed based on the typical values of the diffusion coefficients of the alkali ions in the glass that are in the range of $1 \cdot 10^{-12} cm^2/s < D_2 < 1 \cdot 10^{-10} cm^2/s$, as reported by Patscher and Russel[13] for soda-lime and sodium aluminosilicate glasses. It can be concluded that, in the classical approach of ion exchange for glass in a pure molten nitrate batch, because $D_1 >> D_2$ (there are several order of magnitude), we can reasonably assume:

$$\sqrt{\frac{D_2}{D_1}} \simeq 0, \; \delta_R \simeq 1, \tag{14}$$

Indicating; $C_A(x>0,t) = C_2(x,t)$ and $C_A(x<0,t) = C_1(x,t)$, equation (8) for ions $A$ in the glass results:

$$C_A(x>0,t) = C_{II} - (C_{II} - \chi C_I) \cdot \left[ erfc\left(\frac{x}{2\sqrt{D_2 t}}\right) \right] . \tag{15}$$

Let's introduce the following notation: in (15) $x>0$ represent phase 2 that is glass, while $x<0$ represent phase 1 that is ion reservoir, molten salt bath. The coordinate $x=0$ is the molten salt/glass interface. We indicate with $x=0^-$ the interface side towards molten salt, while with $x=0^+$ towards the glass surface. Considering the second assumption, the *A* ions concentration in the reservoir at the interface with glass (x=0⁻) is not significatively reduced during ion exchange. This assumption means:

$$C_A(0^-,t) \simeq C_A(x<0,0) = C_I. \tag{16}$$

The condition (16) coupled with boundary condition (5) allows to express $\chi C_I$:

$$\chi = \frac{C_A(0^+,t)}{C_A(0^-,t)}; \quad \chi C_I = C_A(0^+,t). \tag{17}$$

Based on (16) and (17) equation (15) results:

$$C_A(x,t) - C_{II} = \left(C_A(0^+,t) - C_{II}\right) \cdot \left[erfc\left(\frac{x}{2\sqrt{D_2 t}}\right)\right] \tag{18}$$

The total concentration of alkali at any time, in the glass, is constant ad it is just the sum of concentrations of ions *A* and ions *B*:

$$C_T(x \geq 0,t) = C_A(x \geq 0,t) + C_B(x \geq 0,t). \tag{19}$$

The additional assumption to be considered is the condition of "almost instantaneous" equilibrium condition at the glass surface (*x=0⁺*). As a consequence of this assumption the concentration of *A* ions in the glass surface during ion exchange is constant and it is the sum of the initial concentration of *A* ions and the maximum allowable amount of *B* ions in the glass that can be exchanged: $\gamma C_B(x>0,0)$, where $\gamma$ is an equilibrium coefficient $0 \leq \gamma \leq 1$ and represents the fraction of *B* ions in the glass that can be exchanged, in equilibrium conditions, with *A* ions coming from the reservoir:

$$C_A(0^+,t) = C_A(x \geq 0,0) + \gamma \cdot C_B(x \geq 0,0) = C_{II} + \gamma \cdot C_B(x \geq 0,0). \tag{20}$$

Equation (20) allows to write equation (18) for the "excess" of A ion concentration in glass:

$$C_A(x,t) - C_{II} = \gamma \cdot C_B(x>0,0) \cdot \left[erfc\left(\frac{x}{2\sqrt{D_2 t}}\right)\right] = Cs_A \cdot \left[erfc\left(\frac{x}{2\sqrt{D_2 t}}\right)\right], \tag{21}$$

where $Cs_A$ is the surface concentration of ions *A* in the glass surface that is assumed to achieve an equilibrium constant value in a time much shorter than the ion exchange total immersion time.

In this way, under the assumptions above outlined, we have justified equation (1). Let's now consider a typical ion exchange process in a soda-lime silicate glass immersed for a defined immersion time in a bath of pure potassium nitrate ($KNO_3$). In this case ion exchange equation (13) reads:

$$K^+ + \bar{Na}^+ \rightleftarrows Na^+ + \bar{K}^+ ; \qquad (22i)$$

$$\left[K^+\right] + \left[Na^+\right] = const. \qquad (22ii)$$

The concentrations ($mol/cm^3$) of $K$ ions in the molten salt bath can be calculated considering the molecular weight of Potassium nitrate and its density at the ion exchange temperature while the concentrations of $K$ and $Na$ ions in the glass can be calculated considering the glass chemical composition, the molecular weight of respective oxides and glass density. Data for the indicated moltens salt and glass are reported in Table IV.

Table IV – Data for Molten salt ($KNO_3$) and Soda-Lime Glass

| Molten Salt – $KNO_3$ | | Glass – Soda – Lime | | |
|---|---|---|---|---|
| $MW(g/mol)$ | 101.03 | Chemical composito (wt %) | | Molecular weight MW (g/mol) |
| IX Temperature $T(°C)$ | 450 | $SiO_2$ | 73 | 60.084 |
| Melting Point $T_{MP}(°C)$ | 337 | $Al_2O_3$ | 1 | 101.961 |
| $\rho_{MP}(g/cm^3)$ | 1.865 | $Na_2O$ | 13 | 61.979 |
| $k(g/cm^3°C)$ | 0.000723 | $K_2O$ | 1 | 94.196 |
| $\rho_T(g/cm^3)$ | 1.783 | $CaO$ | 11 | 56.08 |
| Density versus temperature $\rho_T = \rho_{MP} - k(T - T_{MP})$ | | Others | 1 | - |
| | | Density $\rho_G$ (g/cm3) | 2.5 | - |

Based on data reported in Table IV the calculation of relevant concentrations are performed and reported in Table V. The limiting concentration of Potassium ions in the glass is dictated by the concentration of Sodium, when all Sodium in the glass surface is exchanged by Potassium than the equilibrium coefficient $\gamma=1$.

Table V – Concentrations of exchanging ions in molten salt and glass

| **Concentrations calculation results** | **Values ($mol/cm^3$)** |
|---|---|
| $C_I$ – Concentration of Potassium in Potassium Nitrate molten salt bath | 0.0176385 |
| $C_{II}$ – Initial Concentration of Potassium in Glass | 0.000531 |
| $C(Na)$ – Limit concentration of exchangeable Sodium in Glass | 0.0104874 |
| Coefficient $\chi$ for $\gamma=1 = C(Na)/C_I$ | 0.5946 |

The surface concentration of potassium ions in glass ($Cs$) has been evaluated in Appendix 1:

$$Cs = C_2(x \to 0^+, t) = \frac{\chi\left(C_I\sqrt{D_1} + C_{II}\sqrt{D_2}\right)}{\sqrt{D_1} + \chi\sqrt{D_2}} = \chi C_1(x \to 0^-, t). \quad (23)$$

The factor $\chi$ may be considered a function of $\gamma$ with $\chi_{max}=\chi(\gamma=1)$. Equation (23) can be expressed in terms of the ratio $D_2/D_1$:

$$Cs = \frac{\chi(\gamma)\left(C_I + C_{II}\sqrt{\frac{D_2}{D_1}}\right)}{1 + \chi\sqrt{\frac{D_2}{D_1}}}. \quad (24)$$

Equation (24) represent how surface concentration depends on Kinetics of Potassium ions in the molten salt bath and glass (ratio of diffusion coefficients) and glass surface thermodynamics equilibrium conditions represented by the function $\chi = \chi(\gamma)$. In figure 6 it is reported the relative surface concentration $Crel$ as a function of the square root of the diffusion coefficients ratio where $Crel = Cs/C(Na)$.

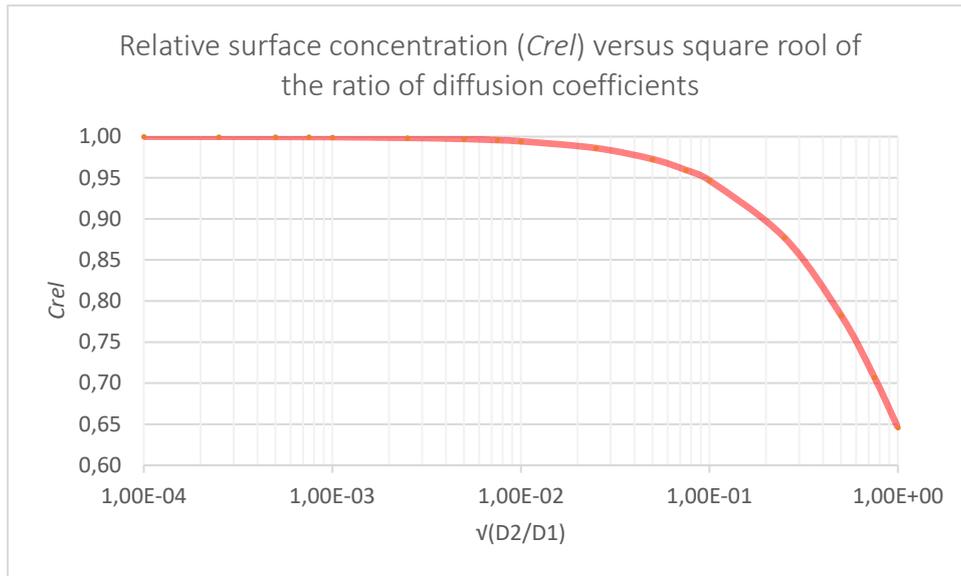

**Figure 6 – Relative surface concentration *Crel* of potassium in glass versus square root of the ratio of diffusion coefficients in glass and molten salt bath ($\gamma=1$).**

From figure 6 it results that the influence of ions kinetics in moltens salt bath generates a reduction to surface concentration in glass when $\sqrt{(D_2/D_1)} > 10^{-2}$. Hence the condition to keep $Crel \approx 1$ is that $D_1 > D_2 \cdot 10^4$. For the higher value of $D_2 = 1 \cdot 10^{-10}$ $cm^2/s$, it results: $D_1 > 10^{-6}$ $cm^2/s$ which is widely satisfied in figure 5 when the exchanging medium is a molten salt bath.

4. **Discussion**

It is important to underline the role of the "equilibrium coefficient" $\gamma$. This has been defined as the fraction of *B* ions in the glass surface that will be exchanged, at equilibrium, with *A* ions coming from the reservoir. On the other side, boundary condition (5) at the interface reservoir/glass (expressed also in equation (17)) connects the interface concentration of *A* ions in the ion reservoir to the surface concentration of *A* ions in the glass. In Figure 7 they are put in evidence the concentration of ions in the ion reservoir (Phase 1) and in the glass before the ion exchange (Straight red line $C_I$ and $C_{II}$ for ions *A* and $C_B(x>0,0)$ for ions *B*). Ideally there is no presence of ions *B* in the reservoir. During ion exchange the assumption of virtually infinite ion reservoir is not affecting $C_I$ curve while at glass surface the equilibrium concentration is:

$$C_{sA} = C_A(0^+, t) = \gamma \cdot C_B(x \geq 0, 0). \tag{25}$$

In Figure 7 it is put in evidence the effect of the $\gamma$-factor in terms of influence to the concentration profile of ions *A* in the glass phase. As pointed out in the initial mathematical study of the general solution (7), the main factor influencing the boundary condition at the reservoir/glass interface are both the diffusion coefficient of ions in the reservoir (phase 1) $D_1$ and the "equilibrium factor" $\gamma$. The applicability of the generally assumed concentration solution (1), based on the constant surface concentration boundary condition, is limited by the underlined assumptions that shall be carefully and critically considered in experimental studies of ion exchange in silicate glasses.

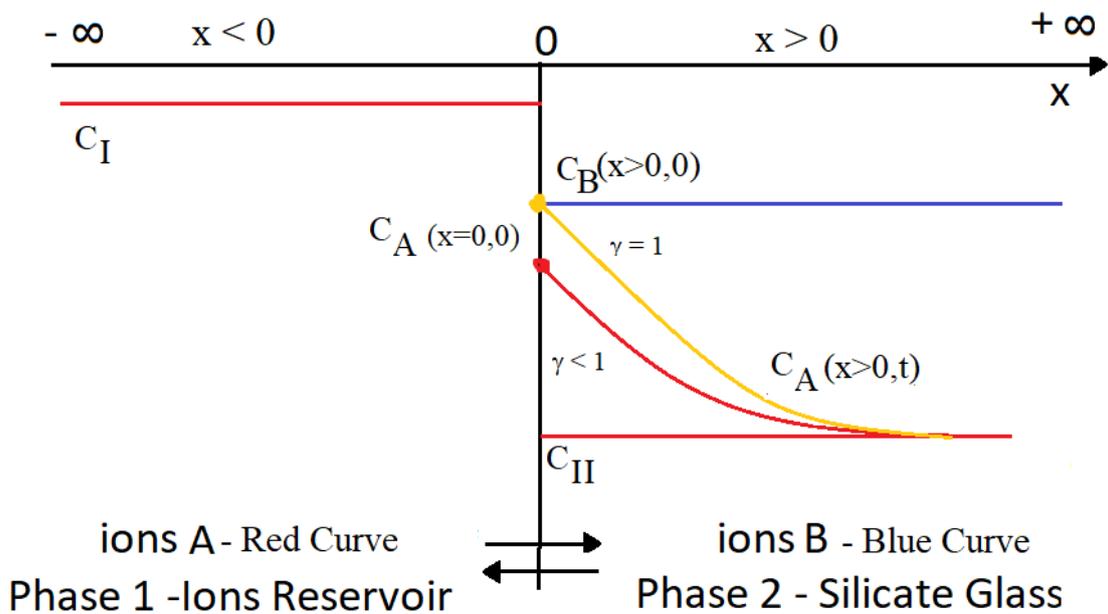

**Figure 7 – Ion Concentration curves, initial conditions ($C_I$ and $C_{II}$), boundary conditions at reservoir/glass interface and effect of $\gamma$-factor on concentration profiles.**

Adding boundary condition (5) to equation (25) allows to express the $\gamma$ coefficient as follows:

$$\gamma = \chi \frac{C_I}{C_B(x \geq 0, 0)}. \tag{26}$$

This has not to be confused with the equilibrium constant that is usually introduced when studying the equilibrium condition for ion exchange [6,13]. Thermodynamic equilibrium condition for the ion exchange process, considered as an isothermal, isochoric process, can be established in terms of the chemical potentials of the exchanging ions[6,14].

$$(\mu_A - \mu_B)_1 = (\mu_A - \mu_B)_2. \tag{27}$$

The thermodynamic equilibrium condition allows to establish an ion exchange isotherm[6] through an equilibrium constant $K$ and a thermodynamic factor $n$. The ion exchange isotherm connects the concentration in the ion reservoir ($C_{1A}, C_{1B}$) with the ones in the glass ($C_{2A}, C_{2B}$):

$$\log\left(\frac{C_{1A}}{C_{1B}}\right) - \frac{W_R}{2.303RT}(1 - 2C_{1B}) = n\log\left(\frac{C_{2A}}{C_{2B}}\right) - \log(K). \tag{28}$$

The important point is that, at the thermodynamic equilibrium, the concentration ratio of the ions in the glass can be correlated to the same concentration ratio in the reservoir:

$$\frac{C_{2A}}{C_{2B}} \approx K \frac{C_{1A}}{C_{1B}}. \tag{29}$$

Equation (29) is relevant when the assumption of "infinite ion reservoir" is no more applicable and a significant concentration of ions $B$ shall be considered in phase 1 (reservoir). The build-up of ions $B$ concentration in the reservoir reduces the equilibrium concentration $C_A(x=0,0)$. The relationship between the equilibrium constant $K$ and the factors $\chi$ and $\gamma$ introduced in this study is not further discussed here. The $\chi$ factor that accounts for the set up of an equilibrium at glass surface when put in contact with the ion reservoir (molten salt bath) shall be related to the thermodynamic equilibrium constant ($K$) but a dedicated investigation is needed to fully clarify the relationship. Another part of the problem is to investigate the mechanism by which equilibrium is achieved that is to clarify the time dependence of the $\chi(t)$ factor. In a previous study[15] a simplified model has been proposed based on the saturation of available sites for ion exchange, that is the original, initial concentration of $B$ alkali ions. Considering the expression of the coefficient $\delta_R$ as a function of the diffusion coefficient $D_1$ of the alkali ions in phase 1 (ion reservoir) a criterium can be derived to approximate equation (12) with equation (1).

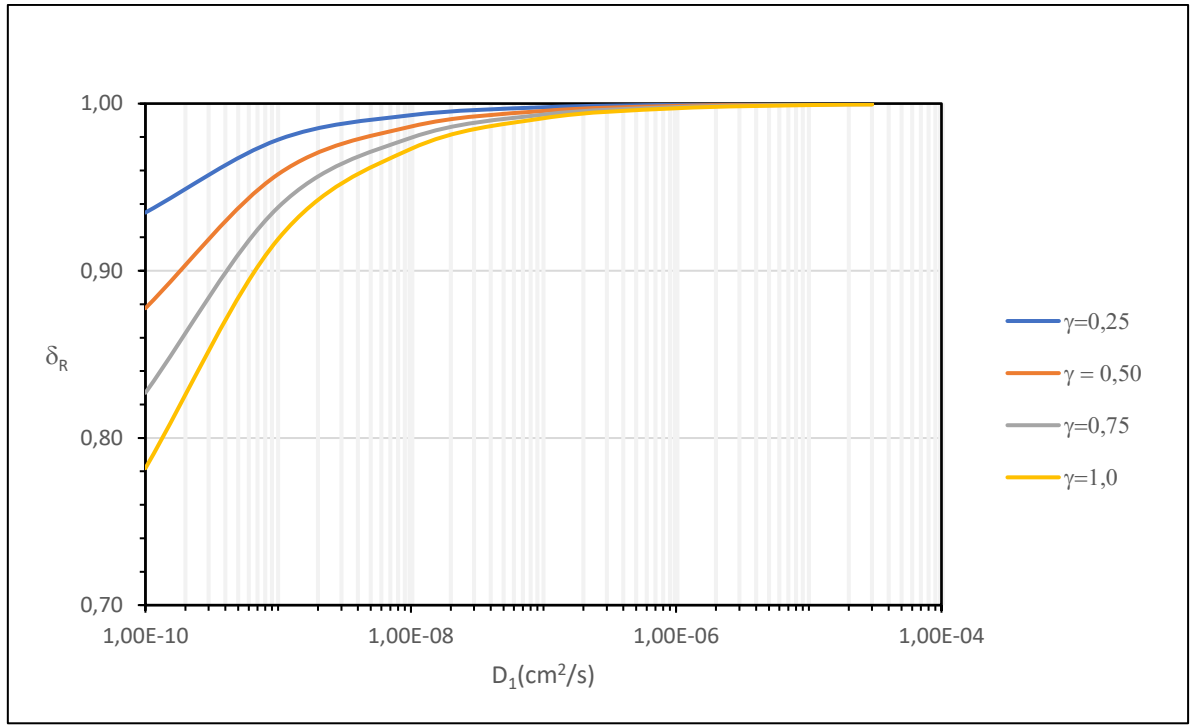

**Figure 8 –** $\delta_R$ **factor (equation (9)) versus Diffusion coefficient of alkali in the reservoir phase. Curves calculated for** $D_2=2.2 \cdot 10^{-11}$ cm$^2$/s **for different values of the equilibrium factor** $\gamma$.

The effect of the equilibrium factor $\gamma$ coupled with the ion kinetics in the molten salt phase can be evaluated by considering the $\delta_R$ factor of equation (9). From figure 8 we can conclude that $\delta_R \approx 1$ when $D_1 > 10^{-6}$ cm$^2$/s, considering the value of $D_2$ in figure 9, it results:

$$\frac{D_2}{D_1} < 10^{-5}. \tag{30}$$

This criterium matches with the conclusion already achieved in figure 6 and it can be translated in a condition for the diffusion length of the ions in the reservoir ($l_R$) and in the glass phase ($l_G$). Diffusion length is proportional to the square root of the product of the diffusion coefficient and diffusion time hence, after equation (30) we have:

$$\frac{l_G}{l_R} = \sqrt{\frac{D_2}{D_1}} < 3 \cdot 10^{-3}, \tag{31}$$

this means that equation (1) is an acceptable approximation providing that the ions diffusion length in the ion reservoir is 300 times larger than the one in the glass phase.

## 5. Conclusion

In this work it has been studied the one-dimensional problem of diffusion of ions in a two phases system where an ion reservoir is put in contact with another phase. Solutions proposed in the literature have been revisited and further clarified with a detailed derivation. They have been clearly identified the assumptions under which the general solution can be reduced to the more familiar complementary error function solution. The theory has been applied to ion exchange in silicate glass and a suitable $\gamma$ - factor has been introduced to extend the equations to this inter-diffusion problem. A criterium has been formulated in terms of ratio of diffusion coefficients or diffusion lengths to accept the approximation of the conventional complementary error function for the ion concentration in glass. Connections have been identified with the equilibrium thermodynamics of ion exchange and further areas of investigation have been identified:

- Correlation of the $\chi$ and $\gamma$ factors to the equilibrium constant of the ion exchange chemical reaction K.
- Investigate the time dependency $\chi(t)$ of the factor connecting the concentration in phase 1 with concentration in phase II.

**Appendix 1.**

Since solution (7) to the two phases one-dimensional diffusion problem has not been justified in details in the literature[7,8,10], herewith it is provided a detailed justification of solution (7). The procedure is straightforward nevertheless there can be some passages requiring attention to get the correct expression of the solution. The introduction of the "ansatz" (6) in the problem defined by equations (2),(3),(4) and (5) allows the determination of the coefficients $E_1$, $E_2$, $F_1$ and $F_2$ as follows:

$$C_1(x,t) = E_1 + F_1 \cdot erf(\frac{x}{2\sqrt{D_1 t}}); \qquad C_2(x,t) = E_2 + F_2 \cdot erf(\frac{x}{2\sqrt{D_2 t}}) \qquad (A1)$$

At t = 0 the argument of the error function in (A1) goes to infinite and it takes values (-∞) when x <0 and (+∞) when x>0 in both cases we have:

$$erf(-\infty) = -erf(\infty) = -1 \qquad (A2)$$

After this result boundary conditions reads:

$$C_I = E_1 - F_1, \quad C_{II} = E_2 + F_2, \quad E_2 = \chi E_1. \qquad (A3)$$

The flux continuity condition at the interface can be calculated considering the derivative of the error function[16]:

$$\frac{d}{dx} erf(z) = \frac{2}{\sqrt{\pi}} e^{-z^2}; \quad -D_1 F_1 \frac{1}{2\sqrt{D_1 t}} = -D_2 F_2 \frac{1}{2\sqrt{D_2 t}}, \quad \frac{F_1}{F_2} = \sqrt{\frac{D_2}{D_1}}, \qquad (A4)$$

The combination of (A3) and (A4) allows to write:

$$E_1 \left( \frac{\sqrt{D_1} + \chi \sqrt{D_2}}{\sqrt{D_1}} \right) = C_I + C_{II} \sqrt{\frac{D_2}{D_1}}; \quad E_1 = \frac{C_I \sqrt{D_1} + C_{II} \sqrt{D_2}}{\sqrt{D_1} + \chi \sqrt{D_2}}, \quad E_2 = \frac{\chi C_I \sqrt{D_1} + \chi C_{II} \sqrt{D_2}}{\sqrt{D_1} + \chi \sqrt{D_2}}, \qquad (A5)$$

From (A3) $F_1$ and $F_2$ can be easily expressed:

$$F_1 = \frac{C_I \sqrt{D_1} + C_{II} \sqrt{D_2}}{\sqrt{D_1} + \chi \sqrt{D_2}} - C_I; \quad F_2 = C_{II} - \frac{\chi C_I \sqrt{D_1} + \chi C_{II} \sqrt{D_2}}{\sqrt{D_1} + \chi \sqrt{D_2}}. \qquad (A6)$$

Based on (A5) and (A6) $C_1(x,t)$ and $C_2(x,t)$ in equations (A1) can be derived:

$$C_1(x,t) = \frac{C_I \sqrt{D_1} + C_{II} \sqrt{D_2}}{\sqrt{D_1} + \chi \sqrt{D_2}} + \left( \frac{C_I \sqrt{D_1} + C_{II} \sqrt{D_2}}{\sqrt{D_1} + \chi \sqrt{D_2}} - C_I \right) erf\left(\frac{x}{2\sqrt{D_1 t}}\right); \qquad (A7i)$$

$$C_2(x,t) = \frac{\chi C_I \sqrt{D_1} + \chi C_{II} \sqrt{D_2}}{\sqrt{D_1} + \chi \sqrt{D_2}} + \left( C_{II} - \frac{\chi C_I \sqrt{D_1} + \chi C_{II} \sqrt{D_2}}{\sqrt{D_1} + \chi \sqrt{D_2}} \right) erf\left( \frac{x}{2\sqrt{D_2 t}} \right). \tag{A7ii}$$

Algebraic manipulations in (A7ii) allows to finally write:

$$C_2(x,t) = C_{II} - \frac{\sqrt{D_1}(C_{II} - \chi C_I)}{\chi \sqrt{D_2} + \sqrt{D_1}} \cdot \left[ 1 - erf\left( \frac{x}{2\sqrt{D_2 t}} \right) \right]. \tag{A8}$$

Recalling that, by definition, *erfc(z)=1-erf(z)* it can be recognized that equation (A8) is exactly the solution indicated by (7) which is, in this way, justified. Equation (A8) can be rewritten in terms of the ratio $D_2/D_1$:

$$C_2(x,t) = C_{II} - \frac{(C_{II} - \chi C_I)}{1 + \chi \sqrt{\frac{D_2}{D_1}}} \cdot \left[ erfc\left( \frac{x}{2\sqrt{D_2 t}} \right) \right]. \tag{A9}$$

In the same way, equation for $C_1(x,t)$, (A7i) can be algebraically further evolved into the following:

$$C_1(x,t) = C_I + \frac{\sqrt{D_2}(C_{II} - \chi C_I)}{\chi \sqrt{D_2} + \sqrt{D_1}} \cdot \left[ 1 + erf\left( \frac{x}{2\sqrt{D_1 t}} \right) \right]. \tag{A10}$$

Equations (A8) and (A10) are the solution to the two phases one-dimensional problem. They can be checked the values at +∞ and -∞ of the two solutions, providing respectively $C_2(x\to+\infty,t)= C_{II}$ and $C_1(x\to-\infty,t)= C_I$. While the values at zero are for the molten salt (ions reservoir):

$$C_1(x \to 0^-, t) = C_I + \frac{\sqrt{D_2}(C_{II} - \chi C_I)}{\chi \sqrt{D_2} + \sqrt{D_1}} = \frac{C_I \sqrt{D_1} + C_{II} \sqrt{D_2}}{\sqrt{D_1} + \chi \sqrt{D_2}} = \frac{C_I + C_{II} \sqrt{\frac{D_2}{D_1}}}{1 + \chi \sqrt{\frac{D_2}{D_1}}}, \tag{A11}$$

and, for the glass surface:

$$C_2(x \to 0^+, t) = C_{II} - \frac{\sqrt{D_1}(C_{II} - \chi C_I)}{\chi \sqrt{D_2} + \sqrt{D_1}} = \frac{\chi\left(C_I \sqrt{D_1} + C_{II} \sqrt{D_2}\right)}{\sqrt{D_1} + \chi \sqrt{D_2}} = \chi C_1(x \to 0^-, t). \tag{A12}$$

Respectively.